\def\bold#1{\setbox0=\hbox{$#1$}%
     \kern-.025em\copy0\kern-\wd0
     \kern.05em\copy0\kern-\wd0
     \kern-.025em\raise.0433em\box0 }
\def\slash#1{\setbox0=\hbox{$#1$}#1\hskip-\wd0\dimen0=5pt\advance
       \dimen0 by-\ht0\advance\dimen0 by\dp0\lower0.5\dimen0\hbox
         to\wd0{\hss\sl/\/\hss}}
\documentstyle[12pt]{article}

\newlength{\dinwidth}
\newlength{\dinmargin}
\newcommand{\resection}[1]{\setcounter{equation}{0}\section{#1}}

\begin{document}
\def\lq{\left [}
\def\rq{\right ]}
\def\LL{{\cal L}}
\def\VV{{\cal V}}
\def\AA{{\cal A}}

\newcommand{\be}{\begin{equation}}
\newcommand{\ee}{\end{equation}}
\newcommand{\bea}{\begin{eqnarray}}
\newcommand{\eea}{\end{eqnarray}}
\newcommand{\nn}{\nonumber}
\newcommand{\dd}{\displaystyle}

\thispagestyle{empty}
\vspace*{4cm}
\begin{center}
  \begin{Large}
  \begin{bf}
    JET ANALYSIS BY NEURAL NETWORKS\\
    IN HIGH ENERGY HADRON-HADRON\\
      COLLISIONS\\
  \end{bf}
  \end{Large}
  \vspace{1cm}
  \begin{large}
P. De Felice, G. Nardulli\\
  \end{large}
{\it I.N.F.N., Sezione di Bari}\\
{\it Dipartimento di Fisica, Universit\'a
di Bari, Italy}\\
  \vspace{8mm}
  \begin{large}
G. Pasquariello\\
  \end{large}
{\it Istituto Elaborazione Segnali Immagini, C.N.R., Bari, Italy}
  \vspace{5mm}
\end{center}
  \vspace{2cm}
\begin{center}
BARI-TH/199-95 \\
January 1995\\
\end{center}
\vspace{1cm}
\begin{quotation}
\begin{center}
  \begin{Large}
  \begin{bf}
  ABSTRACT
  \end{bf}
  \end{Large}
\end{center}
  \vspace{5mm}
\noindent
We study the possibility to employ neural networks to simulate
jet clustering procedures in high energy hadron-hadron collisions. We
concentrate our analysis on the Fermilab Tevatron energy and on the
$k_\bot$ algorithm. We employ both supervised and unsupervised
neural networks. In the first case we consider a
multilayer feed-forward network trained by the backpropagation
algorithm: our results show that these networks can satisfactorily
simulate the relevant features of the $k_\bot$ algorithm. We consider
also unsupervised learning, where the neural network
autonomously organizes the events in clusters. The results of this
analysis are discussed and compared with the supervised approach.
\end{quotation}
\newpage
\setcounter{page}{1}
\resection{Introduction}
\par
Neural Networks (NN) are steadily becoming a standard method of
analysis in high energy physics. Numerical simulations based on the
most common montecarlo codes have been implemented to study
a number of effects
such as discrimination between gluon and quark jets in high energy
$e^+e^-$ collisions \cite{lund1}; $b\overline{b}$ versus light
$q\overline{q}$ production at $Z^0$ peak \cite{alii} \cite{marchesini};
Higgs particle search at future colliders \cite{defelice}\cite{brugnola},
to give only a few examples
(for a review see \cite{bortolotto}). Hardware implementations are becoming
fashionable as well \cite{odorico} and they might offer a clue to
difficult technical problems arising in high energy, high luminosity
future colliders since they might provide
on-line triggers for data acquisition in demanding experimental
environments. In the present letter we wish to address the problem of
the simulation of jet-finding algorithms in high energy hadron hadron
collisions. Intuitively a jet is a collimated {\it spray} of energetic
particles that, when arising from hard parton parton scattering, can
shed light on the short distance QCD dynamics. This intuitive definition
has to be specified for more detailed, quantitative analysis. The
first attempt in this direction
has been represented by the JADE algorithm
\cite{jade} for jet definition in $e^+e^-$ scattering; it introduces a
resolution variable $d^{(J)}_{ij}\,=\,2E_iE_j(1-\cos\theta_{ij})$ for
each pair of particles (jets), having energies $E_i$, $E_j$, with angular
separation $\theta_{ij}$. Once scaled by the total energy:
$y_{ij}\,=\,d^{(J)}_{ij}/Q^2$, this {\it distance} is compared to a
given threshold parameter $y_{cut}$ and the pair belongs to the same
jet provided that $y_{ij}\,\leq \,y_{cut}$.
\par
This first jet definition has evolved into a more sophisticated jet
algorithm, the so-called $k_\bot$ algorithm \cite{kt}, that
we will
briefly review in the next section. The introduction of this
algorithm allows to solve some of the problems found in older
algorithms, such as the attractive kinematic correlation of soft
particles induced by the JADE algorithm or the jets overlap in the Cone
algorithm for hadron hadron scattering \cite{cone} \footnote{For a
nice review of the competitive advantages of the $k_\bot$ algorithm over
JADE or {\it Snowmass} definition of jet, see \cite{catani}.};
moreover the $k_\bot$ clustering algorithm has a cleaner theoretical
foundation \cite{catani} and clear advantages in the small $y_{cut}$
region, since it allows resummation at all orders in $\alpha_s$ of
large double logarithmic corrections arising from soft collinear gluon
emissions.
\par
$k_\bot$ clustering algorithm is in general slow and time consuming
especially in hadron hadron collisions, where one has to separate jets
arising from hard parton scattering from the soft jets associated
to the two initial beams, and for very high energies, because of the
high multiplicity associated to this scattering. Therefore it may be
worthwhile to study the feasibility either to simulate the $k_\bot$
algorithm by a supervised neural network or to implement an
unsupervised NN wich finds its own way to cluster the particles. These
two approaches will be examined in section 3 and 4 and will be applied
to the event-by-event analysis of the number of jet. Finally in
section 5 we draw our conclusions.
\bigskip
\resection{$k_\bot$ clustering algorithm.}
\par
$k_\bot$ clustering algorithm, as applied to $e^+e^-$ collisions,
uses the following resolution variable
\be
d^{(k_\bot)}_{ij}\,=\,2\min\{E_i^2,E_j^2\}(1-\cos\theta_{ij})
\ee
and $y_{ij}\,=\,d^{(k_\bot)}_{ij}/Q^2$, to be compared to the
resolution parameter $y_{cut}$. When applied to hadron hadron
scattering, the algorithm merges a final state particle $i$ into the
jet $j$ or attributes it to the beam remnants (beam jet),
depending on the smaller value between
\be
d_{ij}\,=\,2\min\{E_{Ti}^2,E_{Tj}^2\}
\sqrt{(\eta_i-\eta_j)^2+(\phi_i-\phi_j)^2}
\ee
and
\be
d_{iB}\,=\,E_{Ti}^2 \; .
\ee
Here $E_{Ti}$ is the transverse energy of the $i$-th particle with
respect to the beam direction, $\eta_i\,=\,\ln \tan( \theta/2)$ is its
pseudorapidity, $\phi_i$ is the azimuth angle with respect to the beam
axis. The jet variables are obtained from the jet 4-momentum $p^\mu_J$
which is defined by
\be
p^\mu_J \;=\;\sum p^\mu_i \; ,
\ee
where the sum runs over all the particles in the jet $J$.
\par
In order to separate the beam remnants from the hard parton jets one
usually examines final state particles twice: in the first step a
rather large value of $y_{cut}$  ($y_{cut}\approx 1$) and
$d_{ij}/Q^2_{hard}$, $d_{iB}/Q^2_{hard}$ are compared. $Q_{hard}$ is a
reference mass (typical values of $Q_{hard}$ are around $10^2
\, GeV$; in our case we use $Q_{hard}\sim 55 \, Gev$).
Once the attribution of the soft remnants to the beams is performed, one again
examines the final state by the
algorithm, with different values of the jet resolution parameter
$y_{cut}$ (typical values are $10^{-2}\div 10^{-1}$).
\par
In the following we shall focus our attention to the event-by-event
analysis of $n_{j}$, the number of hard jets and on the
average energies of the different jets. In this paper we have
chosen to work to high, but not very high energies, i.e. we consider
the case of Tevatron at Fermilab ( $\sqrt{s}\;=\;1.8\,TeV$) and we
defer LHC studies to future analyses. The reason for this limitation
is practical. We choose to analyze all the final particles
arising from hard parton scattering; in other words we exclude the beam jets.
Since we use all the particles of the hard jets we are
able to perform more detailed analyses
and to use unprocessed variables.
This implies that we have to consider rather huge neural networks, as it
will be discussed in more detail in the next sections.
At the Tevatron energy, the number of final
particles originated from hard scattering, $n_f$, may be of the order
of $10^2$. We have selected only events with $n_f \; \leq \; 80$, which
represents more than $70\%$ of the total.
\par
Our study is based on simulated events produced by the Herwig
Montecarlo \cite{herwig}. For each event, we take as input $p_x$, $p_y$,
$p_z$ or alternatively ( $E$, $\eta$, $\phi$ ) for each of the $n_f$
final particles.
\bigskip
\resection{Backpropagation feed-forward neural network simulation.}
\par
Our first task is to simulate the $k_\bot$ algorithm by a
feed-forward NN trained by the backpropagation rule \cite{rume}.
Backpropagation networks have been extensively applied to high energy  physics
\cite{bortolotto} and will not be reviewed here. Suffice it to say that we use
a network with 240 input neurons, one hidden layer of 100 neurons and 5
or 7 output units according to the value of $y_{cut}$. The input
neurons $x_i$ are activated by the momenta $p_x$, $p_y$, $p_z$ of all
the final state particles, ordered with energy; if the final state
contains less than 80 particles, the corresponding inputs are put
equal to zero. The momenta $p_{x_i}$, $p_{y_i}$, $p_{z_i}$ are
normalized in the interval $ [ 0,1 ] $. As usual, also the output
neurons $Y_i$ have values in the interval $ [ 0,1 ] $. More precisely,
the output neurons $Y_i$ get the following attribution during the
training: if the event, when
analyzed by the $k_\bot$ algorithm, contains $i$ hard jets,
then we put $Y_i=1$ and $Y_k=0$ for $k\ne i$. In the second phase, the
so called testing phase, $Y_i$ may have any value in
the interval $ [ 0,1 ] $ and it will be given the value
$Y_i=1$ or $Y_k=0$ according to some threshold parameter $T_h$ (see below).
\par
The number of jets depends on the value of $y_{cut}$. The training
set consists of $\sim\,40,000$ events. When
studied by the $k_\bot$ algorithm,  the average value of hard
jets $<n_j>$, for two values of $y_{cut}$, is given by
$<n_j>=2.9$ for $y_{cut}=10^{-2}$ and $<n_j>=2.0$ for $y_{cut}=10^{-1}$.
\par
For $y_{cut}=10^{-2}$ most of the events are concentrated at the value $n_i=3$
(about $40\%$ ), with $\sim20\%$ of 4-jet events and $25\%$ 2-jet
events; moreover we have a few ($\sim 5\%$) events with only 1 jet,
which can be attributed to imperfect balance of the two beam jets.
For $y_{cut}=10^{-1}$, around $75\%$  of the events have 2 hard jets, with
the remaining part almost equally distributed between 1 and 3 jet events.
\par
During the testing phase,
about $3,500$ events, different from those of the
training set,  have been presented to the NN. We
divide the events in classes of assigned number of jets, i.e.
a given event belongs to the class
$\{l\} (l=1,2,...)$ if its particles are clustered in
$l$ jets by the
$k_\bot$ algorithm. For each class $\{l\}$
we can define a purity $p_l$:
\be
 p_l\;=\;\frac {N_l^a}{N_l^a+N^a_{jl}} \hskip 10pt  (j\ne l)
\ee
and an efficiency $\eta_l$:
\be
\eta_l\;=\;\frac {N_l^a}{N_l}
\ee
where $N_l^a$ is the number of events with $l$ hard jets classified as
belonging to the class $\{l\}$ by the
NN, while  $N^a_{jl}$ is the number of
events with $j$ ($j \ne l$) hard jets interpreted as events with $l$
jets and $N_l$ is the total number of events with $l$ hard jets
(accepted or not).
\par
We can vary $p_l$ and $\eta_l$ by modifying an internal
parameter of the network, i.e. the acceptance parameter
$T_h$. It is defined as follows; in the testing phase, the calculated
output for
the neuron $i, \; Y_i$ will be given the value
\bea
Y_i\;=\;1\;\; if \;\; Y_i\;\geq \;1-T_h \nonumber \\
Y_i\;=\;0\;\; if \;\; Y_i\;< \;1-T_h
\eea
Typical results are in Fig.1 for $y_{cut}=10^{-2}$
 for the events with 1 (Fig.1a) and 2 jets (Fig.1b).
 For 3 jets we get purity $p_3\;\approx\;0.54$
with efficiency $\eta_3$ in the range $0.5\div0.7\;$; for 4-jets a purity
of $0.43$ can be obtained with efficiency $\eta_4\,\sim\,0.4$.
For $y_{cut}=10^{-1}$  one gets better results in terms of purity and
efficiency: for example for 1-jet events $p_1=0.98$ with $\eta_1=0.6$
(see Fig.1c) and for 2 jets $p_2\sim 0.91$ at $\eta_2=0.6$ (see
Fig.1d).  We have used in this analysis ( $\vec {p_j}$) as input
variables; had we used ( $\eta_i,\,\phi_i,\,E_i$ )
variables as inputs,
similar results would have been obtained.
\bigskip
\resection{Analysis by an unsupervised competitive NN}
\par
In this section we shall make use of unsupervised competitive learning
\footnote{For an introduction to the subject of unsupervised neural
networks see\cite{hertz}, chap.9.}
to study the feasibility of a neural network that
implements a clustering algorithm without  preliminary supervised
training. Among the various NN approaches using unsupervised training,
here we choose to adopt a self organizing architecture
\footnote {For a more detailed description of the self organizing map
algorithm see\cite{kohonen}; for other applications in high energy
physics see, e. g., \cite{lonn}.}. More precisely,
we use a single layer network with $N=240$ neurons in
the input layer and an output layer of $M$ neurons. The output neurons
can be arranged on a square lattice: we have used $M$ ranging from
 $5^2$ to $20^2$ (better results are obtained with larger values
of $M$). At each time step a new event   $\vec{x}=\{x_i\}
(i=1,\ldots,N)$ is presented as an input to the network and the
distance
\be
d_k\;=\;|\vec{x}-\vec{W_k}|\;=\;
\sqrt{\sum_{i=1}^N(x_i-W_{ik})^2}
\ee
for any output neuron $k$ $(k=1,\ldots,M)$ is computed.  Here $W_{ik}$ is the
element of
the weight
matrix ({\it synaptic matrix}) connecting the input neuron $i$ to the output
neuron $k$; the values of the weight matrix are
chosen initially random and small. Among the output
neurons let $m$ be the one with the smallest distance from $\vec{x}$:
\be
d_m\;\leq\;d_k \;\; \forall\;\; k=1,\ldots,M \; ;
\ee
in this case the output neuron
$m$ becomes the winner and the synapses are modified as
follows:
\bea
W_{ij} \; & \rightarrow  &\; W_{ij} \;+\;\Delta W_{ij} \nonumber \\
\Delta W_{ij} \; & = &\; \eta_{j}\;(x_i \;-\; W_{ij}) \label{updat}
\eea
In the so-called winner-take-all version of the algorithm one puts
$\eta_j=\eta\delta_{jm}$, i.e. only the weights of the winner
neuron $\{W_{im}\}$
are modified; $\eta$ is a positive parameter and the result is to
shift $W_{im}$ towards $x_i$. We have used the self-organizing version
of this algorithm, with $\eta_j=\eta\;\Lambda(j,m)$, where $\Lambda(j,m)$
is a function peaked at $j=m$ and rapidly decreasing with the distance
between $j$ and $m$. This ensure that not only $m$, but also its
neighbours change their weights towards $\vec{x}$. The result of the
updating rule (\ref{updat}) is that, after several iterations, $\vec{W_m}$
yields a representation of all the events that have rendered the
output neuron $m$ the winner. Moreover output neurons that are close
in distance have similar weights.
\par
In our case, in principle, the number of the output neurons $M$ could
be as small as $3^2$,
since the number of jets obtained by the $k_\bot$ analysis never excedes $7$.
In practice, however, more neurons are needed since
the topologies of the events having the same number of jets can widely
differ from each other. Once we rotate the events so that the
most energetic particle is along the positive $z$ axis, this fixes
an average direction for the first jet, but the other ones can be scattered
in any other direction. Therefore more neurons are needed to take into
account the different kinematical configurations. After several
presentations (we have used the same training set of $40,000$ events
employed in the supervised analysis that we have
described previously) one
can adopt two different strategies to analyze the learned weights.
\par
{\bf I)} \hskip 3pt The events are first
analyzed by the $k_\bot$
algorithm; each event after this analysis can be, therefore,
labelled by an integer
number $n_j$, which specifies the jet multiplicity, i.e. the number
of jets in the
event.

Now let us consider the output neuron $m$; let us suppose that it
has been the winner neuron $\omega^{(m)}_1$ times with events having
$n_j=1$ (i.e. with events with $1$ jet), $\omega^{(m)}_2$ times with
events having 2 jets, etc. Let $\bar\omega^{(m)}_l$ be the largest
among the $\omega^{(m)}_j$'s:  $\bar\omega^{(m)}_l \; = \; \omega^{(m)}_l$
such that $\omega^{(m)}_l \geq \omega^{(m)}_j$
for any $j$. We assume a
majority rule, i.e. if $\omega^{(m)}_l$  is the largest
among the $\omega^{(m)}_j$, then the output neuron $m$ is
considered representative of the class $\{l\}$, i.e. it represents
the class of the events having $l$ jets.
We can now define purity ($p_l$) and efficiency ($\eta_l$)
for each class $\{l\}$ of events by formulae analogous to
those of previous section. We define
\be
p_l\;=\;\frac {N_l^a}{N^a_{tot,l}}
\ee
\be
\eta_l\;=\;\frac {N_l^a}{N_l}  \; .
\ee
Now $N_l^a$ has
the following definition:
\be
N_l^a\;\; = \sum_{m|l} \omega^{(m)}_l \; ,
\ee
where the symbol $m|l$ means that the sum runs over
the all the output neurons $m$ which, according
to the majority rule, represent the class $\{l\}$, i.e. are considered
representatives of the events with $l$ jets.
In other terms $N_l^a$ is obtained by summing all the
events with $l$ jets, provided they have been {\it accepted},
which, in this context, means that they have been used to modify
the weights of the output neurons of the class $\{l\}$.
Analogously,
\be
N^a_{tot,l} = \sum_{m|l} \sum_j \omega^{(m)}_j
\ee
represents the total number of {\it accepted} events, i.e. events
that have been attributed to the class $\{l\}$.
Finally, as before, $N_l$, is the total number of events with $l$ jets.
\par
The results we have obtained by this analysis are as follows. First
of all we consider the distribution of the events in the output square
lattice of the $20^2$ neurons. For $y_{cut}=10^{-2}$, it is given by the
Lego plot on Fig.2. We see clearly that 1 jet events are concentrated
in a few neurons at the center of the output square, the 3-jets events
are mostly concentrated at the borders, whereas the neurons
representative of the events with 2 jets  are in an intermediate position.
The diagram in Fig.2 is useful to illustrate the topology of the
output neurons, but is of no use to get quantitative results. They
can be obtained using the previous definitions of purity
and efficiency for the different classes. Some of these results
are reported in Table 1. For each class one can get several results for
the pair (purity, efficiency) by modifying the rule $(4.2)$
as follows:
\be
d_m\;\leq \min \{t,d_k\}  \;\; \forall\;\; k=1,\ldots,M \; ;
\ee
where $t$ is an internal parameter;
if no output neuron satisfies the previous condition the event is
discarded. In Table 1 the two columns are obtained with two different
values of $t : t\; = \; 0.33$ (first column), $t \; = + \infty$ (second
column).

{\bf II)} Unsupervised competitive neural architectures can be used in a
different way; since $\vec{W_k}$ supplies an internal representation
of the patterns that have activated the neuron $k$, we can
interpret, for each output neuron $k$, $\vec{W_k}\;=\;\{W_{ik}\}$
 as the distribution of the
particle momenta of an hypothetical {\it event} that we call call $W_{ik}$
{\it event}.
The events
  represented by $\vec{W_k}$ can be analyzed by the $k_\bot$
algorithm. In other terms we can use the network as a model of the
sample of the physical events that have been used to construct $W_{ik}$.
Since the number of the $W_{ik}$
events is $M\;\leq\;20^2$, much smaller
than the number of events in the original
sample ($\sim 40,000$), it is clear that in this way one can significantly
reduce the time needed for the analysis. It is also evident that, due to
simplifying
assumptions (for example we have discarded events with more than $80$
particles), the results that can be obtained by this method are
approximated; in other terms in real situations this method can be used
as a preliminary classifier of events with a given number of jets;
after this screening of the input data,
 the events of a particular class of interest
might be analyzed by the more precise (but time consuming) $k_\bot$ algorithm.
The results obtained by this analysis are as follows. First of all
one can compute the average number of jets $<n_j>$ using the hypothetical
events $W_{ik}$.
For $y_{cut}\;=\;10^{-2}$, one obtains:
\be
<n_j> \;\;=\;\; 2.84   \hskip 1.cm (\; <n_j> \;\;=\;\; 2.9\;)
\ee
while for $y_{cut}\;=\;10^{-1}$:
\be
<n_j> \;\;=\;\; 1.75   \hskip 1.cm (\; <n_j> \;\;=\;\; 2.0\;)
\ee
where  the values given in parentheses are the results of the $k_\bot$ analysis
on the original $40,000$ events. We can see that the results obtained by
the $k_\bot$ algorithm on the hypothetical $W_{ik}$ events are
similar to those obtained analyzing the full sample of the original events.

A similar analysis can be performed
on the average energies of the jets. Let us consider two groups of events:
group {\it A} (events with particles clustered in 2-jets) and
group {\it B} (events with particles clustered in 3-jets). We compute,
for two values of $y_{cut}: \; y_{cut}=\; 0.1$ and $0.01$ the average
energies of the 2 jets of the group {\it A} and those of the 3
jets in the group {\it B} (the jets are ordered in energy). Again we
perform two computations, one with the {\it physical} events, i.e with
the original $40,000$ events and the other one with the
$W_{ik}$ events. Table 2 shows that the results are rather similar.

\resection{Conclusions}
Our study shows that neural networks can be usefully
employed to simulate
jet clustering procedures, in particular
the $k_\bot$ algorithm,
in high energy hadron-hadron collisions. We
have considered  both supervised and unsupervised
neural networks. In the first case we have used a
multilayered feed-forward network trained by the backpropagation
algorithm and we have shown
that this network can satisfactorily
simulate the average number of jets
as a function of $y_{cut}$.
We have also considered
unsupervised learning, in particular self-organizing competitive
neural networks, characterized by autonomous organization  of
the events in clusters. Our results show
that the clusterization produced by this
network has significant similarities with that induced by the
$k_\bot$ algorithm.

\vskip 1cm

{\bf Acknowledgements.} We wish to thank G. Marchesini for
his collaboration at an early stage of this work and for several
helpful comments; we also grateful to P. Colangelo and F. Fiorani
for useful discussions on the subject of this paper and G. Satalino
for help on using the NN simulator.

\newpage

\begin{table}
\vspace {0.25in}
\begin{centering}
\begin{tabular}{|l|l|c|} \hline \hline

{\em Class of jets}  & $(p_l\;,\;\eta_l)$ & $(p_l\;,\;\eta_l)$ \\ \hline
$\{l\}=1 \;$ (1 jet)   & $(0.65\;,\;0.53)$  & $(0.52\;,\;0.48)$   \\ \hline
$\{l\}=2 \;$ (2 jets)  & $(0.53\;,\;0.16)$  & $(0.46\;,\;0.24)$   \\ \hline
$\{l\}=3 \;$ (3 jets)  & $(0.60\;,\;0.02)$  & $(0.46\;,\;0.87)$   \\ \hline
$\{l\}=4 \;$ (4 jets)  & $(0.67\;,\;0.001)$ & $(0.40\;,\;0.03)$   \\ \hline
\end{tabular}
\vspace {0.25in}
\caption{Purity ($p_l$)and efficiency ($\eta_l$) pairs for
different classes of jets. The first column is obtained with $t=0.33$, the
second column with $t=+\infty$. $t$ is defined in eq. (4.8). }
\label{tab:res1}
\end{centering}
\vspace {0.25in}
\end{table}
\par

\begin{table}
\vspace {0.25in}
\begin{centering}
\begin{tabular}{|r||r|r||r|r|} \hline \hline
    & \multicolumn{2}{c||}{\em physical events}
    & \multicolumn{2}{c|}{ $W_{ik}$ {\em events}} \\ \cline{2-5}
  $y_{cut}$   & 0.1 &0.01 &0.1 &0.01 \\ \hline
A)1-st jet & 174.4 & 152.4 & 172.4  & 170.0 \\ \cline{2-5}
2-nd jet  & 103.4 & 85.8 & 58.7 & 58.9 \\ \cline{2-5}
 & & & &  \\ \hline
B)1-st jet & 167.8 & 149.6 & 145.2  & 141.9 \\ \cline{2-5}
2-nd jet  & 90.8  & 80.3 & 64.9 & 52.8     \\ \cline{2-5}
3-th jet &56.7   & 39.6 & 33.0 & 27.0     \\ \hline
\end{tabular}
\vspace {0.25in}
\caption{Single jet energy average values }
\label{tab:res2}
\end{centering}
\vspace {0.25in}
\end{table}
\par

\newpage

\begin{center}
  \begin{Large}
  \begin{bf}
  Figure Caption
  \end{bf}
  \end{Large}
\end{center}
  \vspace{5mm}
\begin{description}
\item [Fig. 1] The purity $p_l$ versus  efficiency
$\eta_l$ for two different values of $y_{cut}$ and for two values of $l=n_j$
 (number of jets);
 {\bf a}: $l=1\;,\;y_{cut}=10^{-2}\;$;
 {\bf b}: $l=2\;,\;y_{cut}=10^{-2}\;$;
 {\bf c}: $l=1\;,\;y_{cut}=10^{-1}\;$;
 {\bf d}: $l=2\;,\;y_{cut}=10^{-1}\;$.
\item [Fig. 2] Distribution of the output neurons  (unsupervised architecture)
according to their jet class (the identification of jet classes refers to
$k_{\bot}$ algorithm with $y_{cut}=10^{-2}$).
\end{description}

\newpage
\begin{thebibliography}{99}

\bibitem{lund1}L. L\"onnblad, C. Petersen and T.
R\"ognvaldsson, Phys. Rev. Lett. {\bf 65} (1990) 1321.

\bibitem{alii} L. L\"onnblad, C. Petersen and T.
R\"ognvaldsson, Nucl. Phys. {\bf B 349} (1991) 675;
C. Bortolotto, A. De Angelis and L. Lanceri, Nucl
Inst. and Methods {\bf A 306} (1991) 457;
L. Bellantoni et al., Nucl. Inst. and Methods {\bf A 310} (1991) 618.

\bibitem{marchesini} G. Marchesini, G. Nardulli and G. Pasquariello, Nucl.
Phys.
{\bf B 394} (1993) 541.

\bibitem{defelice} P. Chiappetta, P. Colangelo, P. DeFelice, G. Nardulli
and G. Pasquariello, Phys. Let. {\bf B 322} (1994) 219.

\bibitem{brugnola} F. Anselmo et al., Nuovo Cim. {\bf 107 A} (1994) 129.


\bibitem{bortolotto} C. Bortolotto, A. de Angelis, N. D. Groot and J. Seixas,
Int. Journ. of Modern Physics {\bf C 3} (1992) 733.

\bibitem{odorico} P. Mazzanti and R. Odorico, Int. Jour. of
Neural Systems, {\bf 3} (1992) 243.

\bibitem{jade} JADE Coll., W. Bartel et al., Zeit. Phys. {\bf C 33} (1986) 23.

\bibitem{kt} S. Catani, Yu. L. Dokshitzer, M. Olsson, G. Turnock and
B. R. Webber, Phys. Let.  {\bf B 269} (1991) 432;
S. Bethke, Z. Kunszt, D. E. Soper and W. J. Stirling, Nucl. Phys. {\bf b 370}
(1992) 310;
N. Brown and W. J. Stirling, Zeit. Phys. {\bf C 53} (1992) 629.
Yu. L. Dokshitzer and M. Olsson, Nuc. Phys. {\bf B 396} (1993) 137.

\bibitem{cone} J. Huth et al., in Proc. of {\it Research Directions
for the Decay, Snowmass 1990}, ed. E. L. Berger (World Scient. , Singapore)
(1992) 134.

\bibitem{catani} S. Catani, in Proc. of {\it Int. Europhysics Conf. on
High Energy Physics, Marseille 1993}, eds. J. Carr and M. Perrottet
(Ed. Frontieres, France), (1994) 771.

\bibitem{herwig} G. Marchesini and B. R. Webber, Nuc. Phys. {\bf B 310}
(1988) 461.

\bibitem{rume}
D. E. Rumelhart, G. E. Hinton and R. J. Williams, {\it Parallel
Distributed Processing: Explorations in the Microstructure of
Cognition} (MIT Press, Cambridge MA) (1986).

\bibitem{hertz} J. Hertz, A. Krogh and R. G.  Palmer,
{\it Introduction to the Theory of Neural Computation} (Addison-Wesley) 1991.

\bibitem{kohonen} T. Kohonen  {\it Self-Organizing and Associative
Memory} (Springer-Verlag, Berlin) (1989)

\bibitem{lonn} L. L\"onnblad, C. Petersen, H. Pi and T. R\"ognvaldsson,
Comp. Phys. Comm. {\bf 67}, (1991) 193.
\end{thebibliography}
\end{document}